\newif\ifdraft

\documentclass[journal]{IEEEtran}
\usepackage{graphicx}
\usepackage[cmex10]{amsmath}
\usepackage{amssymb}
\usepackage{bm}
\usepackage{array}
\usepackage{afterpage}
\usepackage{cite}
\usepackage{fixltx2e}

\ifx\pdfoutput\undefined \else \usepackage{epstopdf}\fi

\ifdraft
\usepackage{svn}
\SVN $Id: em1d.tex 260 2014-03-25 04:27:31Z shadwick $
\usepackage{color}

\fi

\def\LL{\mathcal{L}}
\def\Lkin{\LL_{\rm part}}
\def\Lint{\LL_{\rm int}}

\def\Lion{\LL_{\rm ion}}
\def\Lfield{\LL_{\rm field}}

\def\V{\varphi}

\def\ma{m}

\def\qe{q}

\def\wa{w_\alpha}

\def\gammaa{\gamma_\alpha}

\def\wp{\omega_p}
\def\wpt{\wp\,t}

\def\qi{q_{\scriptstyle\textsc{I}}}
\def\nion{n^{\scriptstyle\textsc{(Ion)}}}
\def\niont{\widetilde n^{\scriptstyle\textsc{(Ion)}}}

\def\energy{W}

\def\Np{N_p}
\def\Ng{N_g}

\def\sump{\sum_{\alpha=1}^{\Np}}
\def\sumg#1{\sum_{#1=1}^{\Ng}}
\def\sumgi{\sumg i}
\def\sumgij{\sumg{i,j}}

\def\D#1,#2;{D^{\scriptscriptstyle(#1)}_{#2}}

\def\xid{\dot\xi}

\def\xia{\xi^\alpha}
\def\xida{\xid^\alpha}

\def\pia{\pi^\alpha}

\def\xixd{\dot\xi_x}

\def\xixa{\xi_x^\alpha{}}
\def\xixda{\xixd^\alpha{}}

\def\pixd{\dot\pi_x}
\def\pixa{\pi_x^\alpha{}}
\def\pixda{\pixd^\alpha{}}

\def\xizd{\dot\xi_z}

\def\xiza{\xi_z^\alpha{}}
\def\xizda{\xizd^\alpha{}}

\def\pizd{\dot\pi_z}
\def\piza{\pi_z^\alpha{}}
\def\pizda{\pizd^\alpha{}}

\def\etad{\dot\eta}
\def\etaza{\eta_z^\alpha{}}
\def\etaxa{\eta_x^\alpha{}}
\def\etazda{\etad_z^\alpha{}}
\def\etaxda{\etad_x^\alpha{}}

\def\zt{\tilde z}
\def\vxt{\tilde v_x}
\def\vzt{\tilde v_z}

\def\At{\widetilde A}
\def\Vt{\widetilde \V}

\def\argzvv{\zt, \vxt, \vzt}

\def\Ad{\skew{7}\dot A}
\def\Add{\skew{6}\ddot A}

\def\dt{\Delta t}
\def\dz{\Delta z}
\def\dzeta{\Delta\zeta}

\def\meas#1{d#1}

\def\dzvvt{\meas\zt\,\meas\vxt\,\meas\vzt}
\def\intdzvvt{\intD\dzvvt}
\def\intdzvv{\intD{\meas z\,\meas v_x\,\meas v_z}}

\def\intdz{\intD{\meas z}}
\def\intdzeta{\intD{\meas\zeta}}

\def\intD#1{\int\!\! #1\:}

\def\eqref#1{(\ref{#1})}

\allowdisplaybreaks[4]

\begin{document}

\title{Variational Formulation of Macro-Particle Models for Electromagnetic Plasma Simulations}

\author{%
	A.~B.\ Stamm,~\IEEEmembership{Member,~IEEE,}
	B.~A.\ Shadwick, E. G. Evstatiev%
	\thanks{This work was supported in part by the US DoE under contract numbers DE-FG02-08ER55000 and DE-SC0008382 and by US Department of Education grant number P200A090156 (GAANN).}
	\thanks{A.~B.\ Stamm and B.~A.\ Shadwick are with the Department of Physics and Astronomy, University of Nebraska-Lincoln, Lincoln, NE 68588-0299, USA.}%
	\thanks{E.~G.\ Evstatiev is with FAR-TECH, Inc., 10350 Science Center Drive, Bldg.~14, Suite 150, San Diego, CA 92121.}%
	\ifdraft
	\thanks{Version: \SVNId}%
	\fi
}

\maketitle

\begin{abstract}
A variational method is used to derive a self-consistent macro-particle model for relativistic electromagnetic kinetic plasma simulations.  Extending earlier work [E.~G.\ Evstatiev and B.~A.\
Shadwick, J.~Comput.\ Phys., vol.\ 245, pp.\ 376--398, 2013], the discretization of the electromagnetic Low Lagrangian is performed via a reduction of the phase-space distribution function onto a
collection of finite-sized macro-particles of arbitrary shape and discretization of field quantities onto a spatial grid.  This approach may be used with both lab frame coordinates or moving window
coordinates; the latter can greatly improve computational efficiency for studying some types of laser-plasma interactions.  The primary advantage of the variational approach is the preservation of
Lagrangian symmetries, which in our case leads to energy conservation and thus avoids difficulties with grid heating.  Additionally, this approach decouples particle size from grid spacing and relaxes
restrictions on particle shape, leading to low numerical noise.  The variational approach also guarantees consistent approximations in the equations of motion and is amenable to higher order methods in
both space and time.  We restrict our attention to the 1-$\mathbf{\tfrac12}$ dimensional case (one coordinate and two momenta).  Simulations are performed with the new models and demonstrate energy
conservation and low noise.
\end{abstract}


\IEEEpeerreviewmaketitle

\section{Introduction}
\label{intro}

\IEEEPARstart C{omputation} plays an indispensable role in contemporary plasma physics research.  The dominant computational method is the particle-in-cell (PIC) method
\cite{Dawson83a,Hockney88,Birdsall:1991aa}.  The computational efficiency and intuitive nature of the PIC method is largely responsible for this longevity.  The PIC method is ubiquitous and its use
routine owing to the ready availability of powerful computer systems.  The computational demands of the PIC method strongly depend on system size and dimensionality.  One-dimensional simulations can
be readily performed on a modern laptop computer, while three-dimensional studies can require thousands of CPU cores and hundreds of thousands of CPU hours \cite{Cowan:2012ac}.  Despite the popularity
of the PIC method, its theoretical underpinnings have been developed in a largely ad-hoc manner by direct approximation of the equations of motion.  For systems governed by variational principles ---
such as collisionless plasmas --- it is well-known that approximations performed at the level of the equation of motion risk the introduction of anomalous behavior, especially in the system
invariants.  In general this is due to such approximations breaking the link between the resulting equations and the underlying variational principle.  Indeed the PIC method suffers from a number of
unphysical artifacts \cite{Langdon:1970aa,Okuda:1972si,Cormier-Michel:2008bs}.  While in some cases, empirical methods exist to suppress the unwanted behavior, the presence of these artifacts can
greatly complicate interpretation of computational results~\cite{Cormier-Michel:2008bs}.

Recently, a general class of macro-particle methods have been developed \cite{Evstatiev:2013aa,Shadwick:2014aa} using a variational formulation based on Low's Lagrangian \cite{Low:1958aa}.  Significantly, this
approach retains the connection between invariants and symmetries through Noether's theorem \cite{Jose:1998aa}.  One immediate consequence of this connection is the absence of grid-heating in these
models \cite{Evstatiev:2013aa}.  Furthermore, this formulation allows for constructing models of arbitrary spatial and temporal order.  In contrast, the overall accuracy of the usual PIC algorithm is
at most second due to the nature of the force interpolation between the gridded field quantities and the (continuous) particle position.  Again in contrast to the usual PIC algorithm, here the
macro-particle shape is arbitrary; the spatial extent is completely decoupled from both the grid-size and the ``smoothness'' of the shape; smoother particle shapes are not necessarily larger.

Here we extend the original electrostatic analysis \cite{Evstatiev:2013aa} to the simplest relativistic electromagnetic system suitable for the study of laser-plasma interactions, the so-called
1-$\tfrac12$ dimensional case.  We retain a single spatial dimension (the laser propagation direction), $z$, and two particle momenta: one in the direction of the laser polarization, $x$, and other in
the propagation direction.  Taking the vector potential to be $\bm A$, we adopt the gauge fixing condition $\nabla\cdot\bm A = 0$, which, due to our geometry, reduces to $\partial A_z/\partial z = 0$.
In an infinite domain, this implies $A_z=0$.  However, in a bounded domain, this condition allows $A_z = f(t)$, where $f(t)$ is determined by Ampere's law.  For the examples we consider, $A_z$ results
in a small correction to the electric field, which we ignore for simplicity.  (In the 3-D case, choosing a particular gauge can be rather complicated as the typical gauge-fixing conditions lead to 
constrained variations; this is a subject of ongoing research and will be discussed in a future publication.)

Our analysis is carried out with time treated as a continuous variable and thus our equations of motion will be expressed as ordinary differential equations in $t$; it is in this continuous-time
setting that conservations laws (resulting from symmetries in the Lagrangian) hold.  Of course, to perform computations with these models it will be necessary to make time discrete and, generically,
conservation laws will only be preserved asymptotically to some order in the time-step, consistent with the accuracy of the integration method.  There appears to be no impediment to constructing
integrators for our models that respect conservation laws to machine precision, say, using the methods of Ref.~\cite{Shadwick:1999ab}.  Recently, implicit methods have been developed that yield exact
energy conservation in the discrete-time case~\cite{Chen:2011aa,Markidis:2011aa,Lapenta:2011aa}.  While these methods formally exactly conserve energy, in practice the level of energy conservation
achieved is determined by the accuracy with which a large nonlinear system of equation can be solved.  Even when computational limitations preclude energy conservation to machine precision, these
methods are free of grid heating and yield energy behavior superior to the traditional PIC methods.  Here we consider only generic integration methods and examine energy conservation in detail in
Section \ref{energy-conservation}.  Developing integrators to exactly conserve energy for our models is a subject of active research by the authors and will be report upon in due coarse.

In general we adopt the conventions of Ref.~\cite{Evstatiev:2013aa}.  We frame our discussion assuming dynamic electrons and immobile ions; generalization to the multi-species case is entirely
straightforward.  We reduce the distribution function to a collection of macro-particles and, to be concise, we proceed directly to represent the potentials using a spatial grid.  While we only
present a Lagrangian formulation, as in the electrostatic case \cite{Evstatiev:2013aa}, a noncanonical Hamiltonian\cite{Morrison80,Weinstein-Morrison81} formulation is also possible.  We will report
on the full three-dimensional case along with the Hamiltonian formulation in a forthcoming publication.

\section{Reduction to Macro-Particles and Gridded Fields}
\label{Lagrangian-PIC}

It has long been known that the Vlasov equation can be obtained from an action principle \cite{Low:1958aa, Klimontovich:1960aa, Galloway:1971aa}.  Given our geometry, the relativistic version of the
Low Lagrangian \cite{Low:1958aa} takes the form
\begin{equation}
	\begin{aligned}
		\LL =& \intdzvvt f_0(\argzvv)\Biggl[ -\ma c^2 \sqrt{1- \frac{v_x^2}{c^2}-\frac{v_z^2}{c^2}}\\[2pt]
		&\mskip100mu{}- \qe\,\V\left(z,t\right) + \frac{\qe}{c}\,v_x\,A_x\left(z,t\right)\Biggl]\\[4pt]
		&{}+ \frac{1}{8\pi}\intdz\mskip-5mu\left[\frac1{c^2}\left(\frac{\partial A_x}{\partial t}\right)^2 - \left(\frac{\partial A_x}{\partial z}\right)^2 + \left(\frac{\partial\V}{\partial z}\right)^2\right]\\[4pt]
		&{}- \qi\intdz\nion(z)\,\V(z,t),
	\end{aligned}
	\label{Lows_Lagrangian}
\end{equation}
where $z(t;\argzvv)$, $v_x(t;\argzvv)$, and $v_z(t;\argzvv)$ are the electron position and components of velocity having initial conditions $z(0;\argzvv) = \zt$, $v_x(0;\argzvv) = \vxt$ and
$v_z(0;\argzvv) = \vzt$, $f_0(\argzvv)$ is the initial electron phase space distribution, $\V$ is the scalar potential, $q$ and $m$ are the electron charge and mass, respectively, $\qi$ is the ion
charge, $\nion$ is a specified (non-evolving) ion density, and $c$ is the speed of light.  Since the ions are stationary their only contribution to the Lagrangian is their coupling to the
electrostatic potential.  Variations of the action obtained from \eqref{Lows_Lagrangian} with respect to the particle positions yields the usual particle characteristic equations.  Variation with
respect to $\V$ yields Poisson's equation with charge density
\begin{equation}
	q\intdzvv f(z, v_x, v_z, t) + \qi\,\nion \, ,
\end{equation}
while variation with respect to $A_x$ yields Ampere's law with current
\begin{equation}
	q\intdzvv f(z, v_x, v_z, t)\,v_x \, .
\end{equation}
The evolution of the distribution function is obtained from $f(z, v_x, v_z, t) = f_0(\argzvv)$, \textit{i.e.,} using the fact that the distribution function is constant along characteristics.

Following Evstatiev and Shadwick \cite{Evstatiev:2013aa}, we represent the phase space distribution function by a collection of macro-particles
\begin{equation}
	f(z, v_x, v_z, t) = \sump\wa\,f_\alpha,
\end{equation}
where
\begin{equation}
	f_\alpha =S[{ z}-\xiza(t)] \, \delta[{ v_x}-\xixda(t)] \, \delta[{ v_z}-\xizda(t)]\,,
\end{equation}
$\wa$ are constant weights, and the function $S$ is the (fixed) spatial extent of the macro-particle, normalized as
\begin{equation}\label{Shape_norm}
	\intdz S[z-\xiza(t)] = 1.
\end{equation}
Substituting our form of the distribution function into the Lagrangian and utilizing Gardner's re-stacking theorem \cite{Gardner:1963rz}, we obtain a reduced Lagrangian
\begin{equation}
	\LL = \Lkin + \Lint + \Lfield + \Lion\,,\label{L_Cont}
\end{equation}
where
\begin{align}
	\Lkin &= -\ma c^2\sump\wa\,\sqrt{1 - \frac{\xixda^2}{c^2} - \frac{\xizda^2}{c^2}}\,,\label{L_kin}\\[4pt]
	\Lint &= -\qe\sump\wa\!\intdz S(z-\xiza)\left[\V(z,t) - \frac{\xixda}{c}\,A(z,t)\right],\label{coupling}\\
	\Lfield &= \frac{1}{8\pi}\intdz\left(\frac1{c^2}\,\Ad_x^2 + A_x\frac{\partial^2 A_x}{\partial z^2} - \V\,\frac{\partial^2 \V}{\partial z^2}\right),\label{field}\\
\intertext{and}
	\Lion &= -\qi\intdz\nion(z)\,\V(z,t)\,.\label{coupling_ion}
\end{align}
We have integrated by parts in the last two terms of $\Lfield$; as we will see below, the motivation for doing so lies with the finite difference expressions appearing in the discrete form of the
field equations.

We now introduce a fixed (uniform) spatial grid $z_i$ with $i\in[1,\Ng]$ and grid spacing $\dz$ with $\V_i(t)$ and $A_i(t)$ being the numerical approximation of $\V(z_i, t)$ and $A_x(z_i,t)$ respectively.
As the particles positions are not constrained to coincide with the spatial grid, some form of interpolation is required to approximate the potentials between grid-points.  Finite elements
\cite{Becker:1981aa} offer a consistent way to perform such interpolations to any accuracy.  Let $\Psi_i(z)$, $i = 1,\ldots,\Ng$ be finite-element basis of some order.  We interpolate $\V$ and $A_x$
between the grid points by
\begin{equation}
	\V(z,t) = \sumgi\V_i(t)\Psi_i(z) \quad\textrm{and}\quad A_x(z,t) = \sumgi A_i(t)\Psi_i(z)\,.
\end{equation}
Thus 
\begin{align}
	\intdz S(z-\xiza)\,\V(z,t) =& \sumgi\V_i\intdz S(z-\xiza)\,\Psi_i(z) \nonumber\\[4pt]
	=& \sumgi\V_i\,\rho_i(\xiza)\,,
\end{align}
and likewise
\begin{equation}
	\intdz S(z-\xiza)\,A_x(z,t) =  \sumgi A_i\,\rho_i(\xiza)\,,
\end{equation}
where
\begin{equation}
	\rho_i(\xiza) = \intdz S(z-\xiza)\,\Psi_i(z)
	\label{deposition}
\end{equation}
is the effective (projected) shape of the macro-particle.  \{See Table A.1 in Ref.~\cite{Evstatiev:2013aa} for explicit expressions for $\rho_i$ for various shape functions, $S(z)$.\} Assuming the
$\Psi_i(z)$ are constructed from Lagrange polynomials, then $\sumgi \Psi_i(z)$ = 1 and
\begin{equation}
	\sumgi \rho_i(\xia) = \sumgi\intdz S(z-\xiza)\,\Psi_i(z) = \intdz S(z-\xiza) = 1\,.
	\label{charge-norm}
\end{equation}
This means that at any instant the total charge deposited on the grid is $\qe\sump\,\wa$ and the total transverse current is $\qe\sump\wa\,\xixda$ (likewise the total longitudinal current is
$\qe\sump\wa\,\xizda$, but in our geometry, this current does not give rise to electromagnetic fields; its effects are contained within Poisson's equation).  That is, at any instant, all of the charge
and current associated with the macro particles is accounted for on the grid.

The interaction terms, \eqref{coupling} and \eqref{coupling_ion}, can now be written as
\begin{align}
	\Lint &= -\qe\sump\wa\sumgi\left(\V_i - \frac{\xixda}{c}\,A_i\right)\rho_i(\xiza)\\
	\intertext{and}
	\Lion &= -\qi\sumgi\nion_i\V_i\,.
\end{align}
Furthermore, we approximate the field terms, \eqref{field}, by expressing the spatial derivatives using finite differences and replacing the integral by a sum over grid points.  Let $K_{ij}$ be a
finite difference analogue of $\partial^2/\partial z^2$, accurate to some order.  We can then write the field Lagrangian as
\begin{equation}
	\Lfield = \frac{\dz}{8\pi c^2}\sumgi\Ad_i^2  + \frac{\dz}{8\pi}\sumgij\left(A_i\,K_{ij}\,A_j - \V_i K_{ij} \V_j  \right).\label{field_d}
\end{equation}
Note that only the symmetric part of $K_{ij}$ contributes to the Lagrangian.  This has the effect of forcing $K_{ij}$ to correspond to a central difference.  Interestingly, nothing prevents the use of
different finite-difference approximations for the scalar and vector potential terms, thus it is possible to use separate grids for the potentials.  For instance, in the case of under-dense
laser-plasma interactions, the vector and scalar potentials can have very different resolution requirements: the vector potential, representing the laser, demands high resolution, while the scalar
potential, representing the plasma response, can be adequately resolved with lower resolution.  Hence, using separate grids may lead to improved computational performance.

\subsection{Equation of Motion}

The equations of motion are obtained from \eqref{L_Cont} by requiring the corresponding action to be stationary under variations of the particle position and of the potentials.  For the particles, the
usual Euler--Lagrange equations
\begin{equation}
	\frac d{dt}\,\frac{\partial\LL}{\partial\xida_{x,z}} - \frac{\partial\LL}{\partial\xia_{x,z}} = 0\,,\label{EOM_xi_alpha_temp}
\end{equation}
give
\begin{align}
	\pixda&= -\frac\qe c\sumgi\frac{d}{dt}\left[A_i\,\rho_i(\xiza) \right]\label{px-conserve}\\[4pt]
	&= -\qe\sumgi\left[\frac1c\,\Ad_i\,\rho_i(\xiza) + \frac\xizda c\,A_i\,\frac{\partial \rho_i(\xiza)}{\partial \xiza}\right]\label{EOM_pxd}\\[4pt]
	\intertext{and}
	\pizda&= - \qe\sumgi\frac{\partial \rho_i(\xiza)}{\partial \xiza}\left(\V_i - \frac\xixda c\, A_i\right),  \label{EOM_pzd}	
\end{align}
where $\pixa \equiv \ma\,\gammaa\,\xixda$ and $\piza \equiv \ma\,\gammaa\,\xizda$ are the usual relativistic particle momenta with
\begin{equation}
	\gammaa = \sqrt{1 + \frac{\pixa^2}{m^2c^2} + \frac{\piza^2}{m^2c^2}}\,.\label{gammaa}
\end{equation}
Note that $\xixa$ is a cyclic variable, and \eqref{px-conserve} is just a statement of conservation of transverse canonical momentum.  It turns out that the numerical implementation is simpler and the
energy conservation properties (see below) are better if we evolve $\pixa$ according to \eqref{EOM_pxd} in preference to using the conservation law, \eqref{px-conserve}.  These evolution equations
correspond to the Lorentz force, however the discretization has the effect of moving the derivative that would act on the potentials in the continuous case to act instead on the particle shape; in
essence an integration-by-parts is performed behind the scenes.

The Euler--Lagrange equation for the scalar potential is simply $\partial\LL/\partial\V_i= 0$, giving
\begin{equation}
	\sumg jK_{ij} \V_j = -\frac{4\pi}\dz\left[\qe\sump\wa\,\rho_i(\xiza) + \qi\,\nion_i\right] , \label{Poisson}
\end{equation}
which is the discretized form of Poisson's equation. Similarly, the Euler--Lagrange equation for the vector potential is
\begin{equation}
	\frac{d}{dt}\left(\frac{\partial\LL}{\partial\Ad_i} \right) - \frac{\partial\LL}{\partial A_i}= 0\,,
\end{equation}
leading to the wave equation
\begin{equation}
	\Add_i - c^2\sumg j K_{ij}A_j = \frac{4\pi\qe c}{\dz}\sump\wa\,\xixda\,\rho_i(\xiza)\,. \label{WaveEq}
\end{equation}

The consequence of integrating by parts in the field terms in the Lagrangian can now be made clear.  Since the terms for both $A_x$ and $\V$ have the same structure, it suffices to consider only $\V$.
Suppose we had not integrated by parts and had introduced \textit{different} finite-difference representations for each factor of $\partial\V/\partial z$ in the Lagrangian, writing
\begin{equation}
	\frac{1}{8\pi}\intdz\left(\frac{\partial\V}{\partial z}\right)^2 \approx \frac {\dz}{4\pi}\sumg{k,l,m}\tfrac12 \left(\D1,kl;\V_l\,\D2,km;\V_m\right),
\end{equation}
where $\D{1,2},ij;\V_j$ is any finite-difference approximation to $\partial\V/\partial z$ at $z_i$.  Differentiating with respect to $\V_i$ (as is done to obtain the equation of motion) we have
\begin{align}
	\tfrac12 \Big(  \D1,ki;\,\D2,km;\V_m &{}+ \D1,kl;\V_l\D2,ki; \Big) \nonumber\\[4pt]
		&= \tfrac12\left(\D1,ki;\,\D2,kj; + \D2,ki;\D1,kj;\right)\V_j \nonumber\\[4pt]
		&= \tfrac12\left(\D1,;{}^T\D2,; + \D2,;{}^T\D1,;\right)_{ij}\V_j \nonumber\\[4pt]
		&= \widetilde K_{ij}\V_j\,.
\end{align}
Regardless of the details of $\D1,;$ and $\D2,;$, $\widetilde K$ is \textit{symmetric}, \textit{i.e.,} $\widetilde K$ corresponds to some central difference.  Thus whether one integrates by parts in
the Lagrangian or not, the spatial difference operators in the wave equation and Poisson's equation always correspond to some form of central differencing.  Performing the integration-by-parts as we
have done leading up to \eqref{field_d}, allows one to directly specify the difference operator ultimately appearing in the field equations.  This is particularly important with regard to the wave
equation as it is hyperbolic and numerical stability \cite{Thomas:1995aa} will have to be considered.  For example, suppose we take $\D1,; = \D2,; = D$ to correspond to second-order central
differences, $D_{ij} = \left(\delta_{i+1,j} - \delta_{i-1,j}\right)/(2\dz)$, where $\delta_{i,j}$ is the Kronecker delta function, then $\widetilde K$ (up to a sign) corresponds to the standard
second-order central difference for the second derivative but with \textit{twice the grid-spacing}:
\begin{align}
	\tfrac12\Bigl(\D1,ki;\,\D2,kj; &{}+ \D2,ki;\D1,kj;\Bigr) = D_{ki}\,D_{kj}\nonumber\\[4pt]
	=& \frac1{(2\dz)^2}\left(\delta_{k+1,i} - \delta_{k-1,i}\right)\left(\delta_{k+1,j} - \delta_{k-1,j}\right)\nonumber\\[4pt]
	=& \frac1{(2\dz)^2}\left(\delta_{k+1,i}\,\delta_{k+1,j} - \delta_{k+1,i}\,\delta_{k-1,j}\right.\nonumber\\
	&\mskip60mu- \delta_{k-1,i}\,\delta_{k+1,j} + \delta_{k-1,i}\,\delta_{k-1,j})\nonumber\\
	=& -\frac1{(2\dz)^2}\left(\delta_{i+2,j} - 2\,\delta_{i,j} + \delta_{i-2,j}\right).
\end{align}
(Here, the sign change is the same as occurs under integration by parts.)

\subsection{Energy Conservation}

Since our Lagrangian has no explicit time dependence, we will have a conserved energy, $\energy$, which can be obtained from the Lagrangian in the usual way:
\begin{equation}
	\energy = \sump \left(\xixda\,\frac{\partial\LL}{\partial\xixda} +  \xizda\,\frac{\partial \LL}{\partial\xizda}\right) +  \sumgi\Ad_i\,\frac{\partial\LL}{\partial\Ad_i} -\LL\,.
	\label{E0}
\end{equation}
Evaluating $\energy$ with the discretized Lagrangian we obtain
\begin{equation}
	\begin{aligned}
		\energy ={}& \ma c^2\sump\wa\gammaa+ \qe\sump\sumgi\wa\,\V_i\,\rho_i(\xiza)\\[4pt]
		&{} +  \qi\sumgi\nion_i\V_i +  \frac{\dz}{8\pi c^2}\sumgi\Ad_i^2\\[4pt]
		&{}+  \frac{\dz}{8\pi}\sumgij\left(\V_i K_{ij}\V_j  - A_i K_{ij} A_j\right).
	\end{aligned}
\end{equation}
Using the discrete form of Poisson's equation \eqref{Poisson}, we can write the energy in the more recognizable form
\begin{multline}
	\energy =  \ma c^2\sump\wa\,\gammaa + \frac{\dz}{8\pi c^2}\sumgi\Ad_i^2\\ 
	- \frac{\dz}{8\pi}\sumgij\left(\V_i K_{ij} \V_j + A_i K_{ij} A_j\right),
\end{multline}
where the first term is the kinetic energy of the particles and the remaining terms give the discrete representation of the field energy.  Using the equations of motion, it is straightforward to show
that $\energy$ is an invariant:
\begin{align}
	\frac{d\energy}{dt} = {}& \ma c^2\sump\wa\,\frac{d\gammaa}{dt} +  \frac{\dz}{4\pi c^2}\sumgi\Ad_i\,\Add_i  \nonumber\\[4pt]
	&{} - \frac{\dz}{4\pi}\sumgij\left(\V_i K_{ij} \dot\V_j  +  \Ad_i K_{ij} A_j  \right) \nonumber\\[4pt]
	={} & \ma c^2\sump\wa\left(\frac{\partial\gammaa}{\partial\pixa}\,\frac{d\pixa}{dt} + \frac{\partial\gammaa}{\partial\piza}\frac{d\piza}{dt}\right) \nonumber\\[4pt]
	&{}+ \frac{\qe}{c}\sumgi\Ad_i\sump\wa\xixda\,\rho_i - \frac{\dz}{4\pi}\sumgij \V_i K_{ij} \dot \V_j\,,
\end{align}
where we have used \eqref{WaveEq}.  From \eqref{gammaa}, we find $\ma c^2 \partial\gammaa /\partial\pia_{x,z} = \xida_{x,z}$. In addition, we obtain $\dot\V_j$ from the time derivative of \eqref{Poisson}. Together these give
\begin{align}
	 \frac{d\energy}{dt} ={}& \sump\wa\left(\xixda\,\pixda + \xizda\,\pizda\right) \nonumber\\[2pt]
	 &{} + \qe\sump\wa\sumgi\left(\frac{\xixda}c\,\Ad_i\,\rho_i + \V_i \frac{d \rho_i}{dt}\right) \nonumber\\[2pt]
	 ={}& 0\,,
\end{align}
where the last step follows from the macro-particle equations of motion, \eqref{EOM_pxd} and \eqref{EOM_pzd}.

\subsection{Examples}

\begin{figure*}[h!t]
	\centering
	\includegraphics{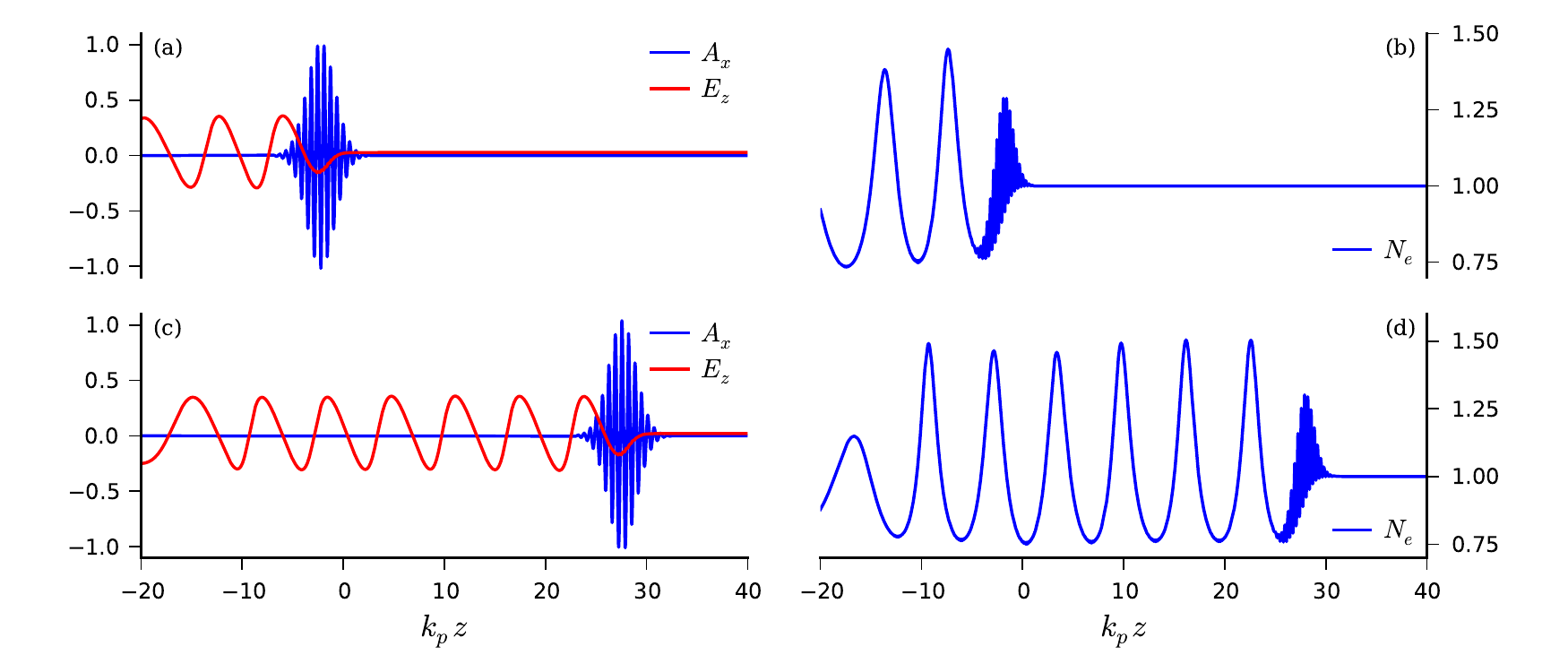}
	\caption{Laser pulse interacting with an under-dense plasma at $\wpt = 50$ [panels (a) and (b)] and $\wpt = 80$ [panels (c) and (d)].  Panels (a) and (c) show $q\,A_x/mc^2$ (red) and
	$q\,E_z/mc\,\wp$ (blue), while panels (b) and (d) show $N_e/n_0$.  \label{under-dense-f}}
\end{figure*}

\begin{figure*}[!ht]
	\centering
	\includegraphics{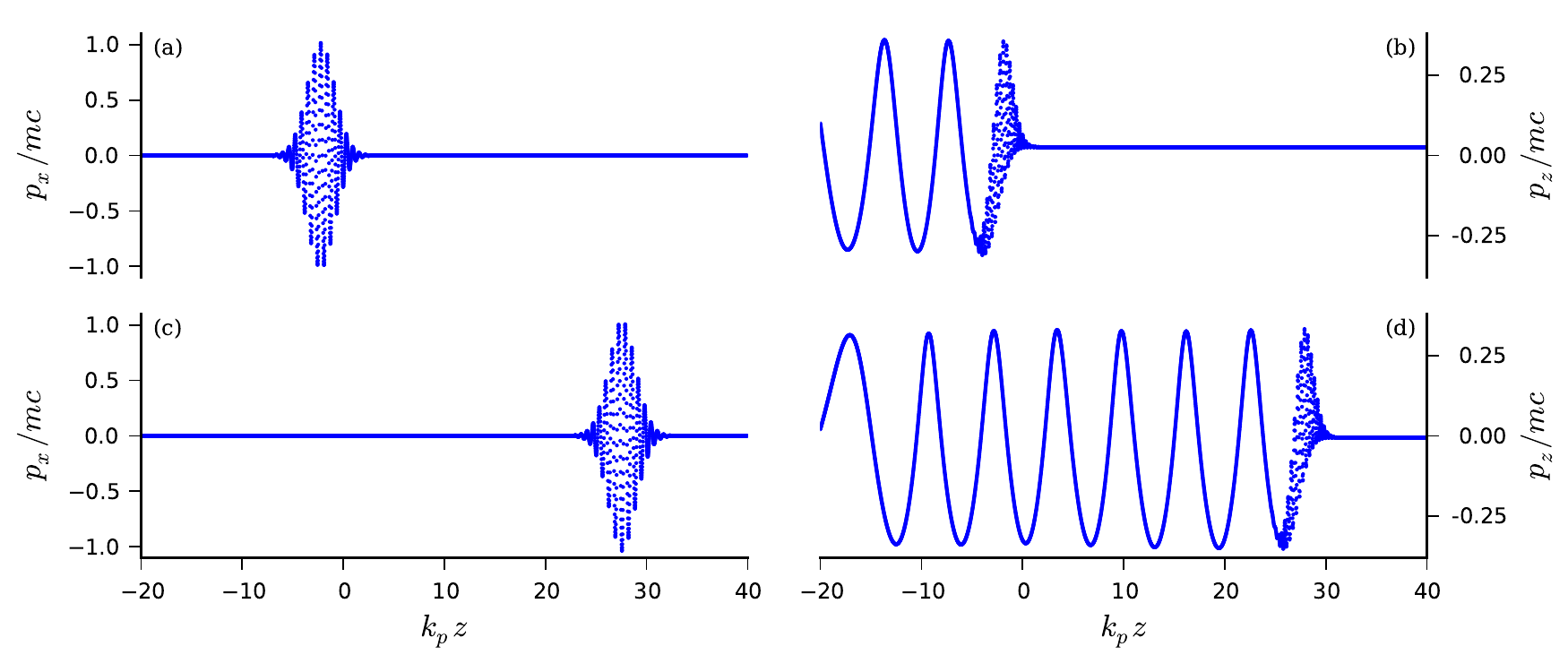}
	\caption{Macro-particle phase space resulting from the interaction of a laser pulse with an under-dense plasma at $\wpt = 50$ [panels (a) and (b)] and $\wpt = 80$ [panels (c) and (d)].  Panels (a)
	and (c) show $\pixa/mc$ and panels (b) and (d) show $\piza/mc$.  Each dot corresponds to a single macro-particle.\label{under-dense-ps}}
\end{figure*}

Throughout we have treated time as a continuous variable, nonetheless, a numerical solution of the equations of motion necessarily requires discretizing in time.  Our formalism is indifferent to the
method used to integrate the equations of motion.  Indeed, it is a significant advantage of our approach that the spatial and temporal discretizations are fully decoupled.  The choice of spatial
differencing (which enters through $K_{ij}$) essentially determines which temporal discretizations will be stable \cite{Thomas:1995aa}.  Thus the absolute freedom to choose the temporal integration
scheme ensures that numerically stable algorithms can be constructed.  We consider two different time integration methods.  For simplicity, in both cases, we adopt second-order finite differences in
space and linear finite elements, $\Psi_i(z)$, for interpolation (also accurate to second-order).  (This interpolation scheme is the same as used in by Evstatiev and Shadwick~\cite{Evstatiev:2013aa}.)
Empirically, we find that numerical stability of this system is dominated by the free space behavior of the wave equation.  A simple stability analysis of the wave equation shows that, with
second-order spatial differencing, second order, explicit time integration is unconditionally unstable~\cite{JPR-2013}.  While both third and fourth order methods are stable, the stability limit for
fourth order is larger~\cite{JPR-2013}.  This leads us to choose a fixed-step fourth-order Runge--Kutta scheme \cite{Butcher:1987aa}; in the following, we will refer to this as ``RK4.''

Alternatively, the Crank--Nicholson scheme \cite{Crank:1947aa} is unconditionally stable for the wave equation.  To avoid a fully implicit solution of the equations of motion, we use Strang-style
operator splitting \cite{Strang:1968aa}, solving the particle equations (including Poisson's equation) with fixed fields and the wave equation with fixed current.  We use a second order Runge--Kutta
method for the particles (we could have just as well used Milne's method \cite{Abramowitz65}, but it requires more intermediate storage) and the Crank--Nicolson method for the fields.  (The
Crank--Nicolson method for the wave equation is also implicit but leads to a bi-tri-diagonal system of linear equations for which a fast direct method exists~\cite{vonRosenberg:1969aa}.)  Since the
field solve is much less computational effort than the particle advance, we choose to perform a half time-step field solve, followed by a full time-step evolution of the particles and electrostatic
potential and a final half time-step field solve.  Subsequently, we refer to this method as ``RK2-Split.''

Our examples consist of a laser pulse incident on an initially quiescent plasma slab.  We consider two cases: $\omega_0 = 10\,\wp$ (the under-dense case) and $\omega_0 = \wp$ (the over-dense case),
where $\omega_0$ is the initial laser frequency and $\wp = \sqrt{4\pi\,q^2n_0/m}$ is the plasma frequency with $n_0$ the ambient plasma density.  The initial vector potential is given by
\begin{equation}
	A_x = a_0\,\frac{mc^2}q\exp\left[-\left(\frac{z - z_0}L\right)^2\right]\cos\left[k_0\left(z - z_0\right)\right],
	\label{Ax-initial}
\end{equation}
where $k_0 = \omega_0/c$ is the initial laser wave number, $z_0$ is the initial location of the center of the pulse and $L$ is the pulse length.  Initially, $\partial A_x/\partial t$ is chosen to
correspond to a forward propagating pulse.  We impose conducting boundary conditions, taking both $\V = 0$ and $A_x = 0$ at the boundary.  The computational grid extends from $z_1$ to $z_2$ and
corresponds to the interior of the problem domain, i.e., the boundary condition are applied at $z_1 - \dz$ and $z_2 + \dz$.  The ion density profile varies from vacuum to a uniform plateau of density
$n_0$ as a linear ramp with quadratically rounded corners.  At the center of the transition, $z_r$, the ramp has slope $2n_0/L_r$; the entire transition has length $L_r$.  Macro-particles are loaded
at rest with variable weights to give a charge-neutral initial density.  All computations are done in dimensionless form with length and time-scales determined by $k_p = \wp /c$ and $\wp$
respectively; momenta are normalized to $mc$, and potentials to $mc^2/q$.

\subsubsection{The Under-Dense Case}
Here we take $\omega_0 = 10\,\wp$, $a_0 = 1$, $k_p\,L = 2$, $k_p\,z_0 = -50$, $k_p\, z_1 = -60$, $k_p\,z_2 = 90$, $k_p\,L_r = 40$, $k_p\,z_r = -30$, and we use one macro-particle per cell.  The long
ramp was chosen to minimize particle trapping at the vacuum-plasma interface.  This problem is solved over a range of grid parameters and with both the RK2-Split and RK4 methods; see below.  Figures
\ref{under-dense-f} and \ref{under-dense-ps} show results with the highest resolution ($k_p\,\dz = 0.05$, corresponding to 12001 grid points, and $c\,\dt=\dz/8$) and quartic $\rho_k$ using the RK4
method at $\wpt = 50$ and $\wpt = 80$.  (See Table A.1 of Ref.~\cite{Evstatiev:2013aa} for explicit expressions for the particle shapes and the $\rho_k$.)  In Fig.~\ref{under-dense-f}, we plot the
dimensionless fields $q\,A_x/mc^2$, $q\,E_z/m\,c\,\wp$ [panels (a) and (c)], and $N_e/n_0$ [panels (b) and (d)].  We compute the longitudinal electric field, $E_z$, from the potential
\begin{equation}
	E_z(z_i) = \frac1{2\dz}\left(\V_{i-1} - \V_{i+1}\right)
\end{equation}
and define the macro-particle density on the grid, $N_e$, based on the right hand side
of Poisson's equation:
\begin{equation}
	N_e(z_i) = \sump\wa\,\rho_i(\xiza)\,.
\end{equation}
In Fig.~\ref{under-dense-ps}, we plot the dimensionless macro-particle momentum $\pixa/mc$ [panels (a) and (c)] and $\piza$ [panels (b) and (d)].

\begin{figure}[!t]
	\centering
	\includegraphics[width=252pt]{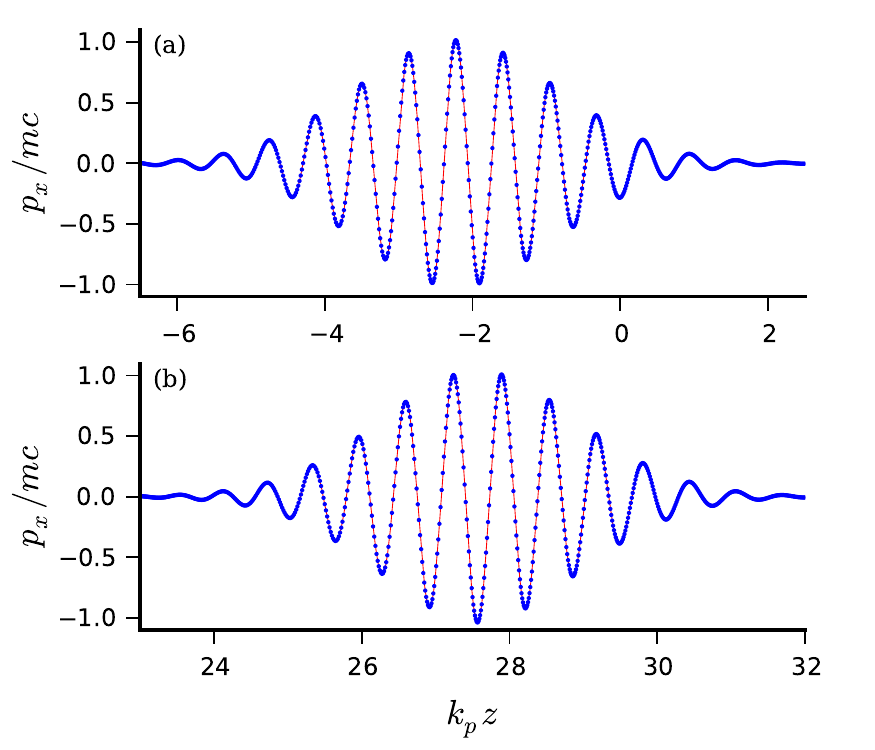}
	\vskip-0.125in
	\caption{Comparison of the macro-particle transverse momentum $\pixa/mc$ (blue dots) and $-q\,A_x/mc^2$ (red line) at $\wpt = 50$ [panel (a)] and $\wpt = 80$ [panel (b)].\label{under-dense-cm}}
\end{figure}

As can be seen in the figures, a clean and well-defined plasma wave is generated.  It should be emphasized that neither the fields (including the density) nor the phase space have been smoothed in any
way.  As mentioned above, we choose to evolve $\pixa$ using \eqref{EOM_pxd} in place of the conservation law, \eqref{px-conserve}.  Figure~\ref{under-dense-cm} shows the transverse momentum overlaid on
$-q\,A_x/c$ at $\omega_p t = 50$ and $\omega_p t = 80$.  The spatial grid is sufficiently fine that $A_x$ is nearly constant over the macro-particle, leading to $\pixa\approx -q\,A_x/c$ to a very good
approximation, as can be seen in Fig.~\ref{under-dense-cm}.

\subsubsection{The Over-Dense Case}
Here we take $\omega_0 = \wp$, $a_0 = 0.5$, $k_p\,L = 10$, $k_p\,z_0 = -40$, $k_p\, z_1 = -75$, $k_p\,z_2 = 75$, $k_p\,L_r = 15$, $k_p\,z_r = 0$, and we use 10 macro-particles per cell.  This problem
is solved over a range of grid parameters and with both the RK2-Split and RK4 methods; see below.  Figure~\ref{over-dense} shows results with the highest resolution ($k_p\,\dz = 0.025$, corresponding
to 6001 grid points, and $c\,\dt=\dz/9$) and quartic $\rho_k$ using the RK4 method.  Plotted in Fig.~\ref{over-dense} are $q\,A_x/mc^2$ and $J_x/q\,n_0$ on the left axis and $N_e/n_0$ on the right
axis at $\wpt=0$ [panel (a)], $\wpt= 50$ [panel (b)], and $\wpt=100$ [panel (c)].  We define the transverse current on the grid, $J_x$, based on the right hand side of the wave equation~\eqref{WaveEq}
as
\begin{equation}
	J_x(z_i) = \frac\qe\dz\sump\wa\,\xixda\,\rho_i(\xiza)\,.
	\label{Jx}
\end{equation}
As can be seen in the figure, the laser pulse is absorbed on the density transition, resulting in ``surface'' currents in the transition region [see Fig.~\ref{over-dense}(b)], which subsequently
re-radiate a left-going pulse as well as an evanescent wave [see Fig.~\ref{over-dense}(c)].

\begin{figure}[!t]
	\centering
	\includegraphics[width=252pt]{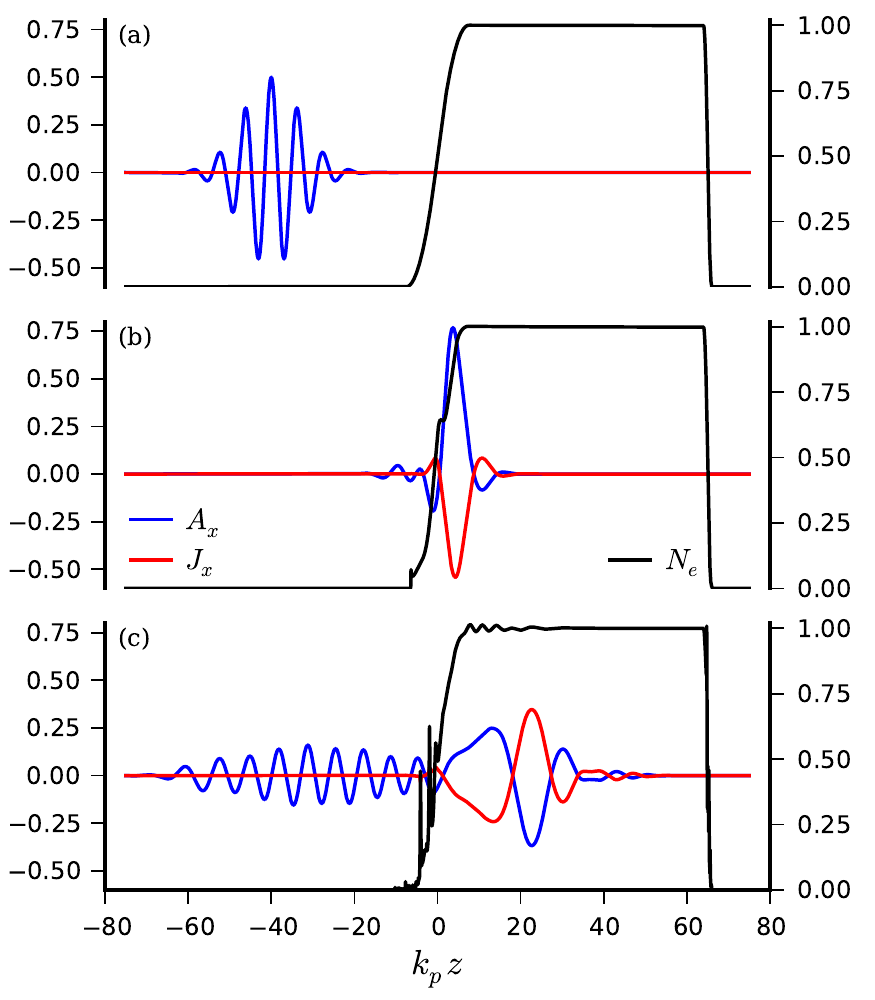}
	\caption{Laser pulse interacting with an over-dense plasma at: (a) $\wpt = 0$; (b) $\wpt = 50$; and (c) $\wpt = 100$.  The vector potential $q\,A_x/mc^2$ (red line) and transverse current
	$J_x/q\,n_0$ (blue line) are plotted on the left axis, while the macro-particle density $N_e/n_0$ (black line) is plotted on the right axis.  The density shows the vacuum-plasma
	interface.\label{over-dense}}
\end{figure}

\begin{figure*}[p]
	\centering
	\includegraphics{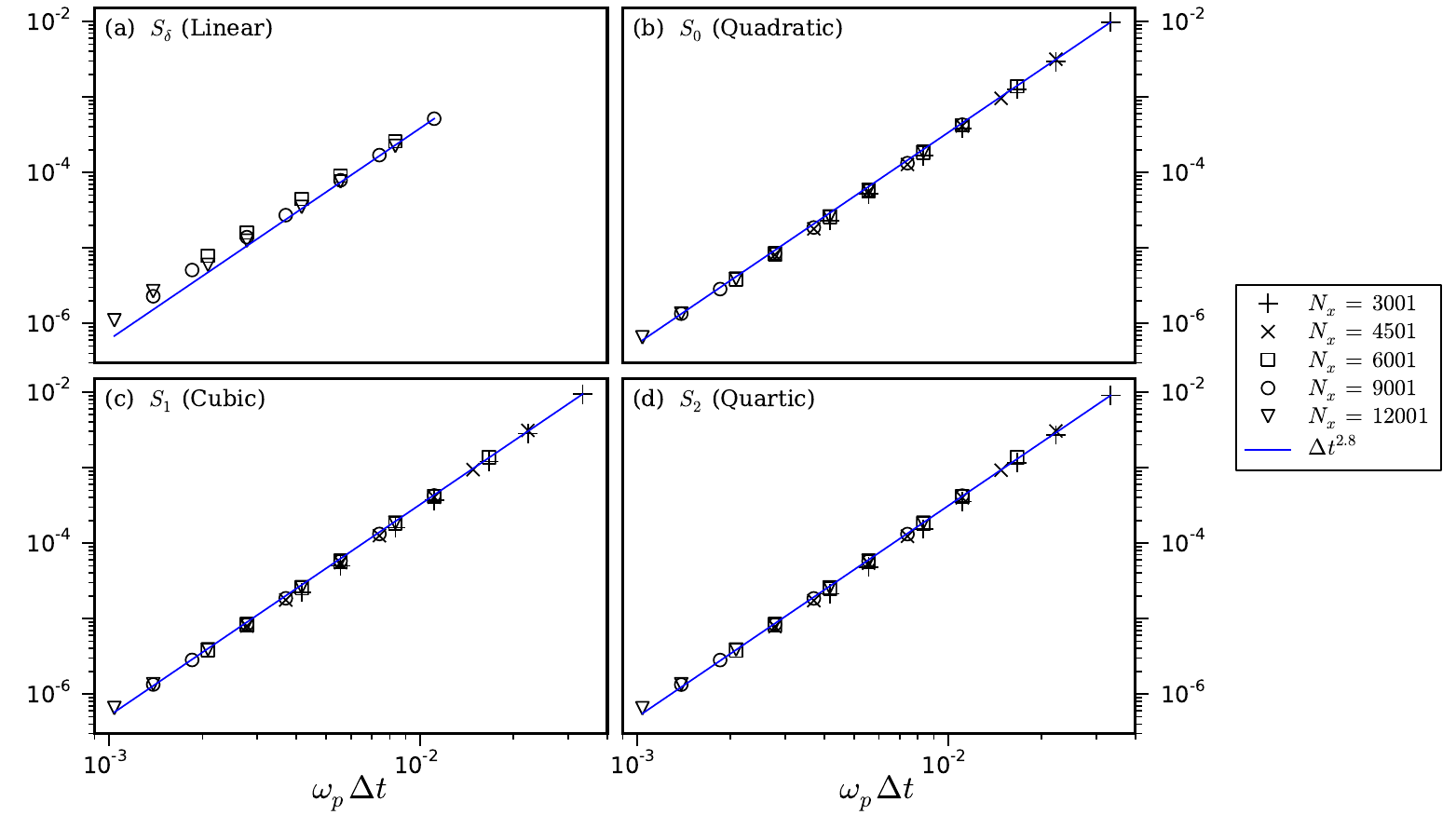}
	\caption{Energy conservation in the under-dense case for the RK2-Split method.  The relative energy error is shown as a function of $\dt$ for various spatial resolutions and particle shapes.  As
	expected, the energy error depends only on the temporal discretization.  The panels are labeled with the particle shape $S$ and the resulting order of $\rho_{k}$.  The shapes are named following
	Ref.~\cite{Evstatiev:2013aa}.\label{energy-rk2-u}}
	\vskip0.25in%
	\includegraphics{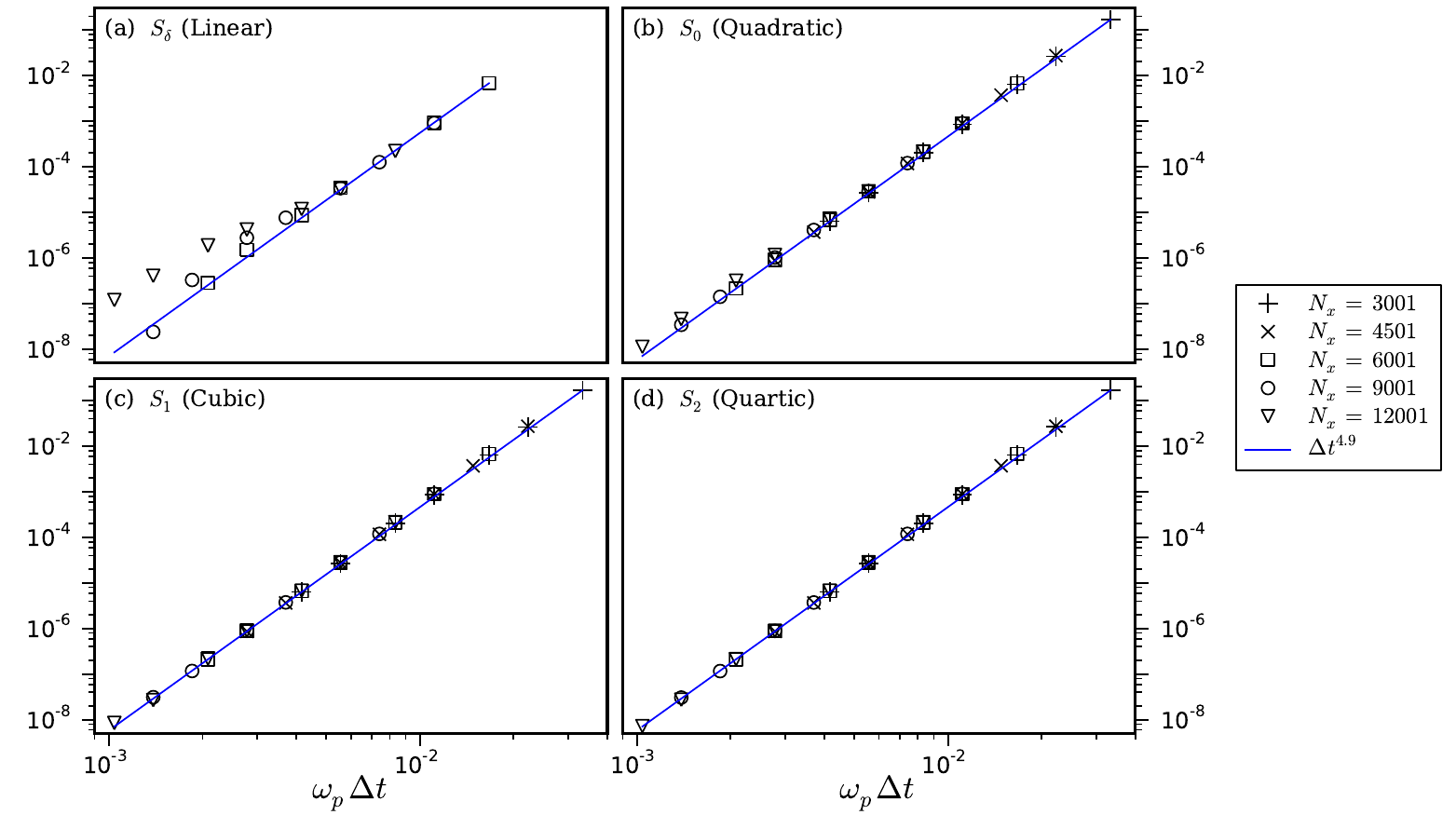}
	\caption{Energy conservation in the under-dense case for the RK4 method.  The relative energy error is shown as a function of $\dt$ for various spatial resolutions and particle shapes.  As
	expected, the energy error depends only on the temporal discretization.  The panels are labeled with the particle shape $S$ and the resulting order of $\rho_{k}$.  The shapes are named following
	Ref.~\cite{Evstatiev:2013aa}.\label{energy-rk4-u}}
\end{figure*}

\begin{figure*}[p]
	\centering
	\includegraphics{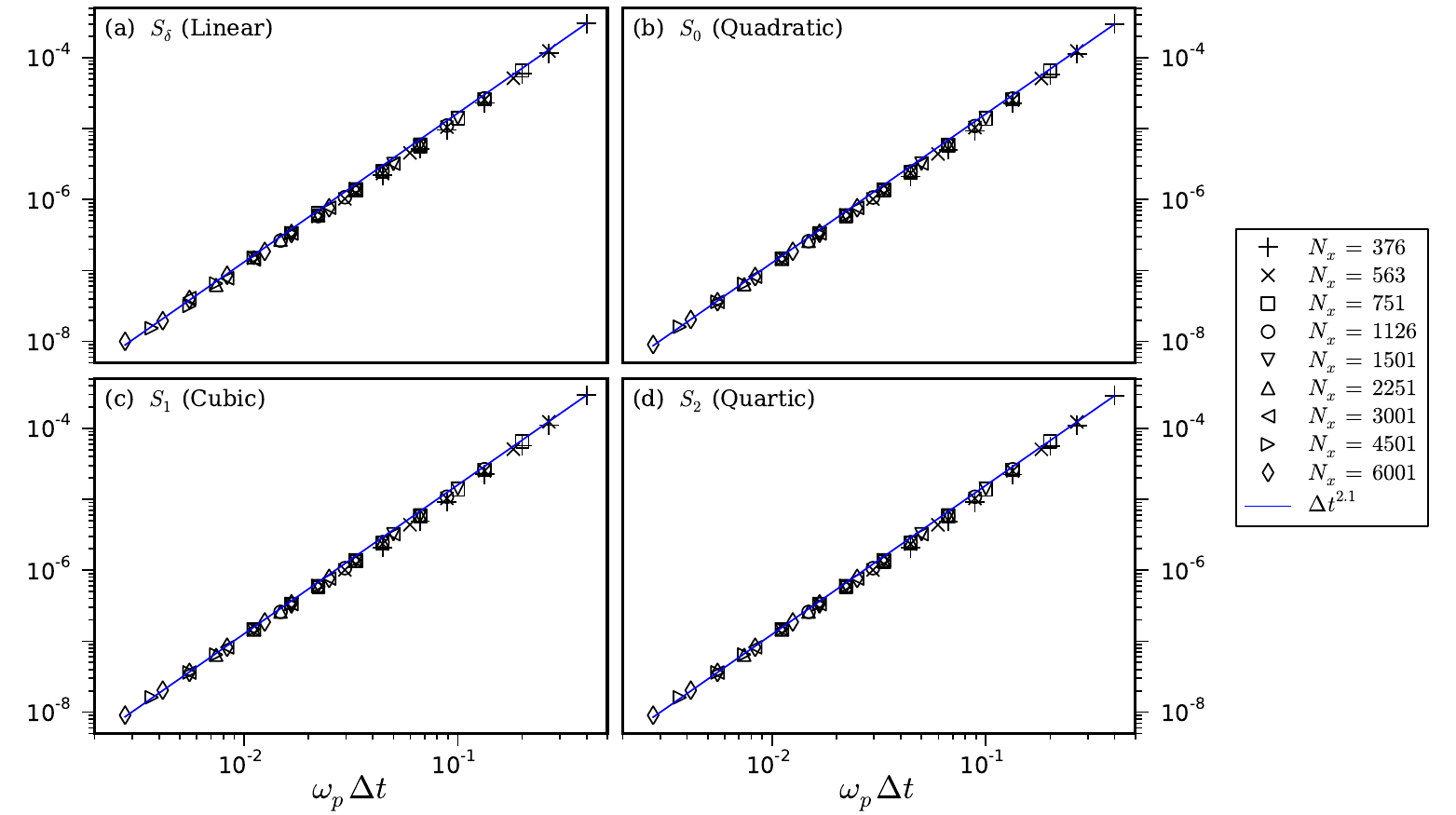}
	\caption{Energy conservation in the over-dense case for the RK2-Split method.  The relative energy error is shown as a function of $\dt$ for various spatial resolutions and particle shapes.  As
	expected, the energy error depends only on the temporal discretization.  The panels are labeled with the particle shape $S$ and the resulting order of $\rho_{k}$.  The shapes are named following
	Ref.~\cite{Evstatiev:2013aa}.\label{energy-rk2-o}}
	\vskip0.25in%
	\includegraphics{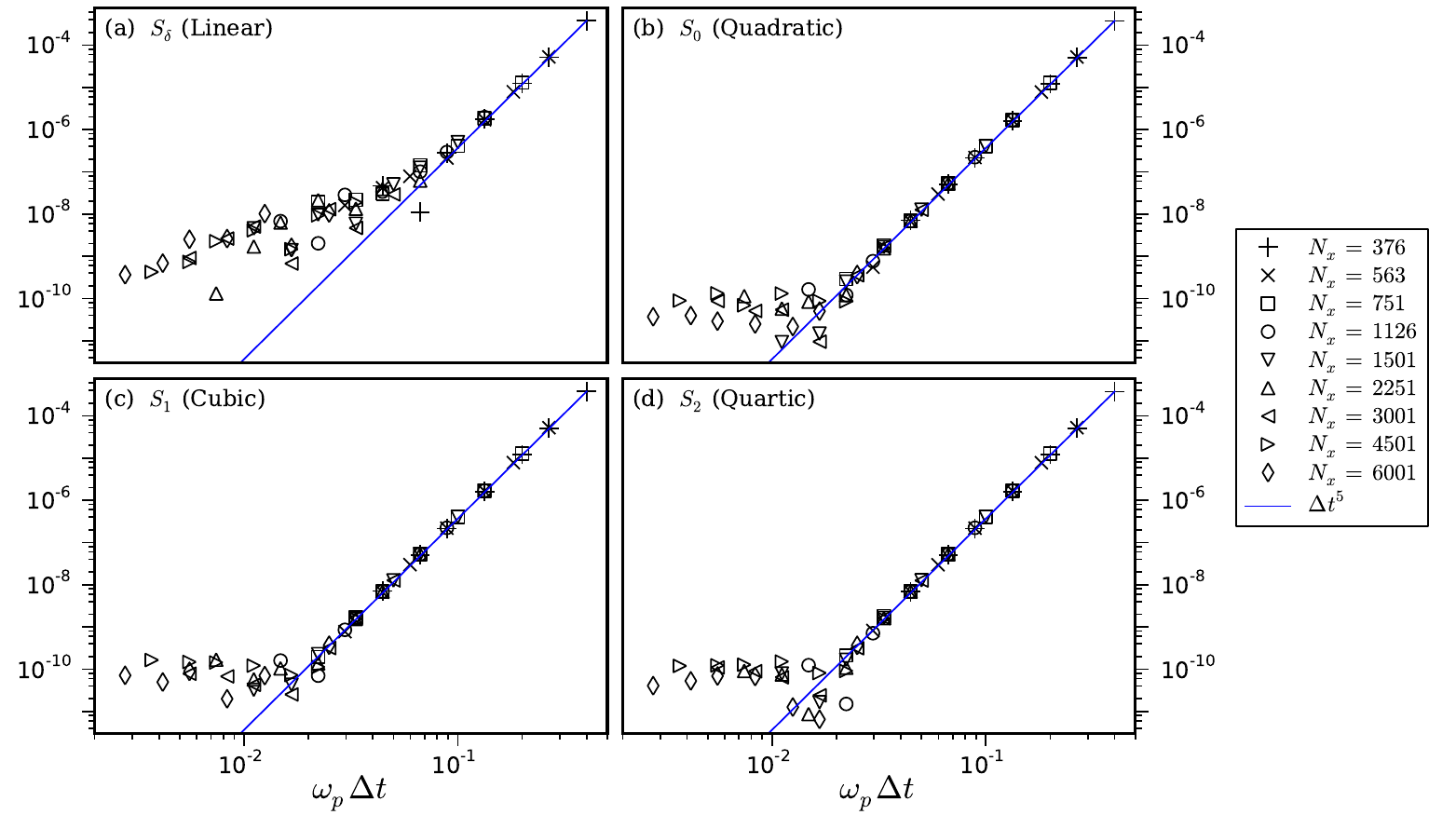}
	\caption{Energy conservation in the over-dense case for the RK4 method.  The relative energy error is shown as a function of $\dt$ for various spatial resolutions and particle shapes.  As
	expected, the energy error depends only on the temporal discretization.  The panels are labeled with the particle shape $S$ and the resulting order of $\rho_{k}$.  The shapes are named following
	Ref.~\cite{Evstatiev:2013aa}.\label{energy-rk4-o}}
\end{figure*}

\subsubsection{Energy Conservation}\label{energy-conservation}
As we saw, the continuous-time equations of motion exactly conserve total energy.  When these equations are integrated numerically, we expect, as a consequence of the time-discretization, that energy
will no longer be exactly conserved.  (It may be possible to construct special purpose integrators for these equations of motion that do exactly conserve energy \cite{Shadwick:1999ab}.)  However,
since any departure from exact energy conservation is due solely to the temporal discretization, the resulting error in total energy should then only depend on $\dt$ and should scale with $\dt$
consistent with the order of accuracy of the temporal integration.  This is in marked contrast to the usual PIC algorithm where the energy error in general depends on both the time-step and the grid
spacing.

To demonstrate this characteristic of the energy error we solve both the under-dense and over-dense problems with each method for a collection of grid-sizes and time-steps.  Since the RK4 method has a
stability limit, we set the largest time-step considered to $c\,\dt = \dz$.  (The actual stability threshold is $c\,\dt \le \sqrt{2}\,\dz$\cite{JPR-2013}.)  The RK2-Split method has a large
stability basin, however, for $c\,\dt > \dz$, there is substantial dispersion in the field solver, thus for accuracy reasons we restrict $c\,\dt \le \dz$ for this method as well.

In Figs.~\ref{energy-rk2-u} and \ref{energy-rk4-u} we plot the relative energy error in the under-dense case for the RK2-Split and RK4 methods, respectively, with $k_p\,\dz$ ranging from $0.05$ to
$0.0125$ and $c\,\dt = \dz$ to $c\,\dt = \dz/8$ for four particle shapes.  In each figure a scaling with $\dt$ is plotted to aid the eye (blue line); the exponent is obtained by fitting the errors.
Panel (a) of Figs.~\ref{energy-rk2-u} and \ref{energy-rk4-u} has fewer points than the other panels due to a technical detail of our implementations.  To simplify our numerical implementations, we
assume no macro-particles leave the domain.  If a macro-particle reaches the domain boundary, the computation is terminated.  At lower resolution, several of the computations with linear $\rho_k$
failed for this reason and are thus absence from the plots.  For the RK2-Split method, we expect the energy error to scale with $\dt^2$ (consistent with the global error of the method).  For all
$\rho_k$ except the linear case, we see nearly perfect power law scaling with $\dt$, with a rather larger exponent than expected.  For linear $\rho_k$, the energy error shows some spread amongst the
different spatial resolutions.  Now, $\partial\rho_k/\partial t \propto\partial\rho_k/\partial\xi$ which, for linear particles, has a discontinuity whose size depends on $\dz$.  As a result, the usual
truncation error analysis does not hold (the numerical method is sampling this derivative and is sensitive to the discontinuity).  For the RK4 method we again see some spread for linear $\rho_k$ and
perhaps some (much smaller) spread for quadratic $\rho_k$.  The quadratic $\rho_k$ have discontinuities in their second derivative which the RK4 method samples.  (We expect to see this also in the
cubic $\rho_k$ but evidently the effect is too small to be observable.)

In Figs.~\ref{energy-rk2-o} and \ref{energy-rk4-o}, we plot the relative energy error in the over-dense case for the RK2-Split and RK4 methods, respectively, with $k_p\,\dz$ ranging from $0.4$ to
$0.025$ and $c\,\dt = \dz$ to $c\,\dt = \dz/9$ for four particle shapes.  In each figure a scaling with $\dt$ is plotted to aid the eye (blue line); the exponent is obtained by fitting the errors.
Overall the behavior is comparable to the under-dense case.  For the RK4 method the departure from the power-law scaling for linear $\rho_k$ is more pronounced than in the under-dense case (whereas
this departure is barely noticeable for the RK2-Split method).  The scatter seen for $\wp\dt \lesssim 2\times10^{-2}$ is due to numerical precision.  While all computations are done in double
precision ($\approx15$ digits), results are stored to disk in single precision ($\approx8$ digits).  For $\wp\dt$ below this threshold, the stored solutions do not have sufficient precision to
faithfully represent the system energy.

In no case do we see any hint of grid heating; this is completely consistent with our formulation which exactly conserves energy even with the presence of a spatial grid.

\section{Moving Window Formulation}
\label{Moving_Win}

A tremendous advantage of the Lagrangian formalism is the Euler--Lagrange equations are form-invariant under arbitrary (invertible) point transformations of the dynamical variables.  For some types of
laser-plasma interactions, moving window coordinates (co-moving with the laser pulse) can greatly reduce the computational cost of simulations.  Here we transform our macro-particle model to moving
coordinates.  While it might be more elegant to apply the transformation to the discrete systems, this is undesirable due to the time-dependence in the transformation.  Thus, we transform the
continuous space macro-particle Lagrangian \eqref{L_Cont} to the moving window coordinates and then discretize the fields.

In the moving window our new coordinates are $\zeta$ and $\tau$ defined by $\zeta = c\,t - z$, $\tau = t$.  Partial derivatives in the two coordinate systems are related by $\partial/\partial z = -
\partial/\partial\zeta$ and $\partial/\partial t = \partial/\partial\tau + c\,\partial/\partial\zeta$.  The new particle positions and velocities become $\etaza = c\,t - \xiza$, $\etaxda \equiv
d\etaxa/d\tau = \xixda$, and $\etazda \equiv d\etaza/d\tau = c - \xizda$.  Under this transformation, the Lagrangian becomes
\begin{equation}
	\LL = \Lkin + \Lint + \Lfield + \Lion\,,
\end{equation}
where
\begin{align}
	 \Lkin &= -\ma c^2 \sump\wa\,\sqrt{1 - \frac{\etaxda^2}{c^2} - \left(1 - \frac\etazda c\right)^2}\,,\label{L-p-mw}\\ 
	 \Lint &= -\qe\sump\wa\!\!\intdzeta S(\etaza-\zeta)\left[\Vt(\zeta,\tau) - \frac{\etaxda}{c}\,\At_x(\zeta,\tau)\right], \label{L-int-mw}\\
	 \Lfield &= \frac{1}{8\pi}\intdzeta\!\!\left[\frac1{c^2}\left(\frac{\partial \At_x}{\partial\tau}\right)^2 + \frac2c\,\frac{\partial \At_x}{\partial\tau}\,\frac{\partial \At_x}{\partial \zeta} - 
	\Vt\,\frac{\partial^2 \Vt}{\partial \zeta^2}\right],\label{L-f-mw}\\
	 \Lion &= -\qi\intdzeta\niont(\zeta, \tau)\,\Vt(\zeta,\tau)\,,\label{L-i-mw}
\end{align}
$\Vt(\zeta,\tau) = \V(z,t)$, $\At_x(\zeta, \tau) = A_x(z,t)$, and $\niont(\zeta, \tau) = \nion(z,t)$.  Note that spatial variation in the ion density leads to time-dependence of $\niont$ in the moving
window. 

We discretize \eqref{L-p-mw}--\eqref{L-i-mw} by introducing a uniform grid $\zeta_i$, $i \in [1,\Ng]$, with spacing $\dzeta$ and follow the procedure described in Section~\ref{Lagrangian-PIC}.  If the
shape function $S(z)$ is symmetric, i.e. if $S(-z) = S(z)$ (there seems little motivation for $S$ to be otherwise), then the projected particle shape, $\rho_k$, in the moving window is identical
to that in the lab frame.  The discrete analogues of \eqref{L-p-mw}--\eqref{L-i-mw} are found to be 
\begin{align}
	\Lkin &= -\ma c^2 \sump\wa\,\sqrt{1 - \frac{\etaxda^2}{c^2} - \left(1 - \frac\etazda c\right)^2}\,,\\[2pt]
	\Lint &= -\qe \sump\wa\sumgi\left(\V_i - \frac{\etaxda}{c} A_i \right) \rho_i(\etaza)\,,  \\[4pt]
	\Lfield &= \frac{\dzeta}{8\pi c^2}\sumgi\Ad_i^2  +  \frac{\dzeta}{4\pi c}\sumgij\Ad_i D_{ij} A_j\nonumber\\[4pt]
	&\mskip160mu- \frac{\dzeta}{8\pi}\sumgij\V_i K_{ij} \V_j\,,\label{field_d-mw}\\
\intertext{and}
	\Lion &= -\qi \sumgi \nion_i \V_i\,,
\end{align}
where $\V_i(\tau)$ and $A_i(\tau)$ are the numerical approximations to $\Vt(\zeta_i,\tau)$ and $\At_x(\zeta_i, \tau)$ respectively.

The equations of motion are obtained in the usual way, giving
\begin{gather}
	\pixda = -\frac{\qe}c\sum_k\frac{d}{d\tau}\left[A_k\,\rho_k(\etaza)\right], \label{EOM_pxd_mw} \\[4pt]
	\pizda = \qe\sum_k\frac{\partial\rho_k(\etaza)}{\partial\etaza}\left(\V_k - \frac{\etaxda}c\,A_k\right),  \label{EOM_pzd_mw}\\[4pt]
 	\sumg j K_{ij}\V_j = -\frac{4\pi}\dzeta\left[\qe\sump\wa\,\rho_i(\etaza) + \qi\,\nion_i\right]\label{Poisson_mw} \\
\intertext{and}
 	\Add_i + c\sumg j \left(D_{ij} - D_{ji}\right)\Ad_j = \frac{4\pi\qe c}{\dzeta} \sump\wa\,\etaxda\,\rho_i(\etaza)\,. \label{WaveEq_mw}
\end{gather}
where $\pixa \equiv \ma\,\gammaa\,\etaxda$ and $\piza \equiv \ma\,\gammaa(c - \etazda)$, with $\gammaa$ given by \eqref{gammaa}.  Note that $\pixa$ and $\piza$ are identical the to corresponding
lab-frame quantities.  Once again the spatial differencing operators are naturally combined in such a way as to corresponding to central differencing.

In an infinite domain, even with a non-uniform ion density, an invariant energy integral can be constructed.  In a bounded domain, since the $\zeta$ domain is moving through space, energy balance 
necessarily requires accounting for particle and field fluxed entering and leaving the domain.

\subsection{Examples}

\begin{figure}[!t]
	\centering
	\includegraphics[width=252pt]{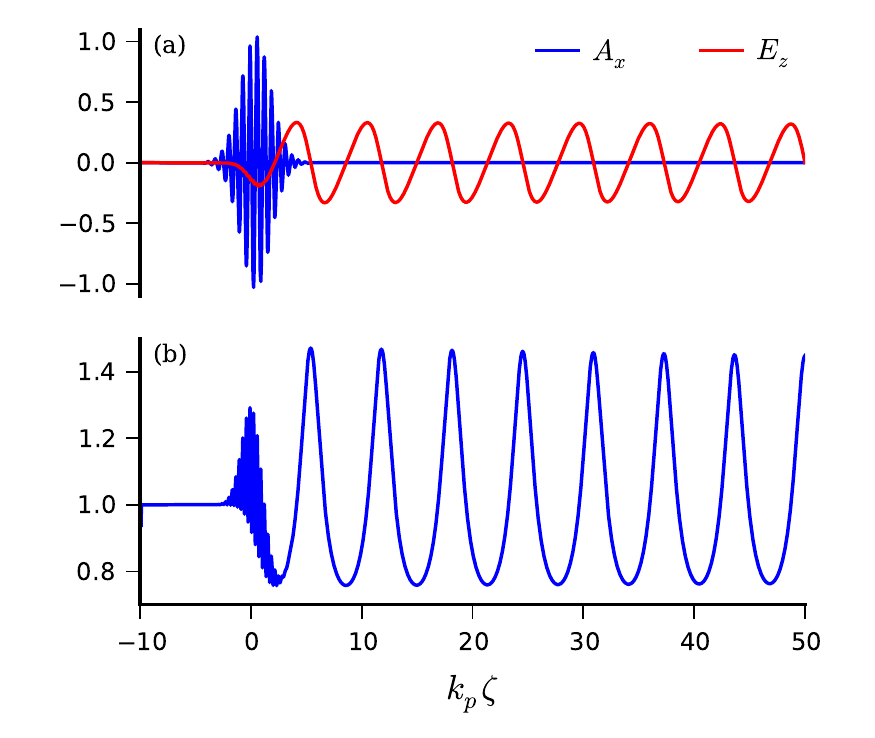}
	\caption{Laser interacting with under-dense plasma in the moving window at $\wpt = 60$.  Panel (a) shows $q\,A_x/mc^2$ (red) and $q\,E_z/mc\,\wp$ (blue), while panel (b) shows $N_e/n_0$.
	\label{moving-window-f}}
\end{figure}

\begin{figure}[!ht]
	\centering
	\includegraphics[width=252pt]{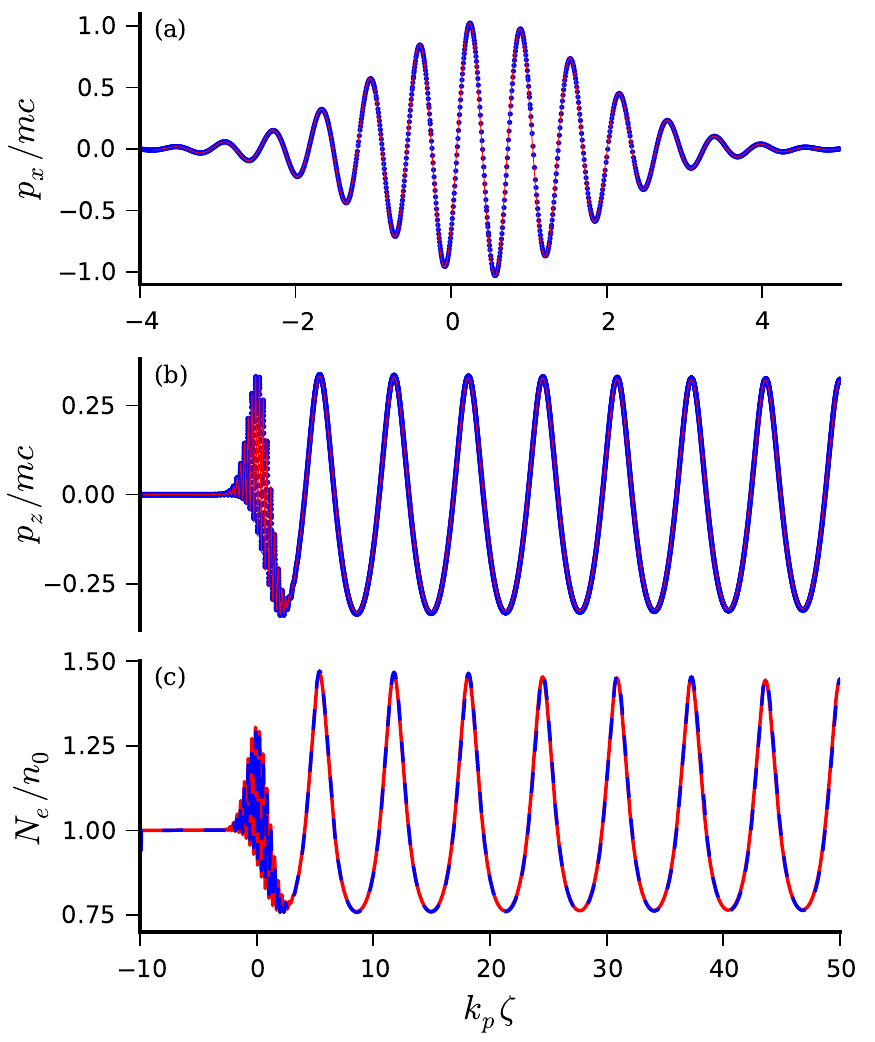}
	\caption{Comparison of our macro-particle model (blue) to the cold fluid model (red) of laser interacting with under-dense plasma in moving window coordinates.  Panels (a) and (b) show phase space
	at $\omega_p t=60$ and panel (c) shows the normalized particle density, $N_e/n_0$.  In panel (a) we only show the area of non-zero $x$-momentum.  \label{moving-window-ps}}
\end{figure}

As in Section \ref{Lagrangian-PIC}, we take second-order spatial differencing and use linear finite-elements for interpolation.  Then $D_{ij} = (\delta_{i+1,j} - \delta_{i-1,j})/(2\dzeta)$ and $D_{ji} 
= - D_{ij}$ and \eqref{WaveEq_mw} becomes
\begin{equation}
	\Add_i + \frac c\dzeta\left(\Ad_{i+1} - \Ad_{i-1}\right) = \frac{4\pi\qe c}{\dzeta} \sump\wa\,\etaxda\,\rho_i(\etaza)\,.
\end{equation}
Again, second-order integrators are unstable \cite{Reyes:2010aa} and we choose to implement a fourth-order Runge--Kutta scheme.  We consider an under-dense plasma with $\omega_0 = 10\,\wp$.  The
initial vector potential is
\begin{equation}
	A_x = \frac{mc^2}q\,a_0\exp\left(-\frac{\zeta^2}{L^2}\right)\cos(k_0\,\zeta),
	\label{Ax-initial-window}
\end{equation}
with $k_p\,L = 2$ (linear resonance) and $a_0 = 1$.  We take $\partial A_x/\partial\tau = 0$, which correspond to forward pulse propagation.  Our boundary conditions are applied ahead of the laser
pulse, i.e., the leading edge of the moving window encounters quiescent plasma, where we take the potentials and their derivatives to be zero.  Our domain extends from $k_p\,\zeta_1 = -10$ to
$k_p\,\zeta_2 = 70$ with 3201 grid-points ($k_p\,\dzeta = 0.025$).  We take $c\Delta\tau=\dzeta$ and use 8 particles per cell.  We use the $S_2$ particle shape, which gives quartic $\rho_k$ (see Table
A.1 in Ref.~\cite{Evstatiev:2013aa}).  In Fig.~\ref{moving-window-f}, we plot the dimensionless fields $q\,A_x/mc^2$, $q\,E_z/m\,c\,\wp$ [panel (a)], and $N_e/n_0$ [panel (b)] at $\wpt = 60$.

Figure~\ref{moving-window-ps} shows a comparison between our macro-particle calculation and the results of a cold fluid model.  The fluid model, also formulated in the moving window, uses the same
spatial differencing, time-integration, grid-parameters and initial conditions.  Figure~\ref{moving-window-ps}(a) shows the macro-particle momentum $\pixa$ (blue dots) overlaid on the transverse fluid
momentum (red line).  Likewise, Fig.~\ref{moving-window-ps}(b) shows the macro-particle momentum $\piza$ (blue dots) overlaid on the longitudinal fluid momentum (red line).  Finally,
Fig.~\ref{moving-window-ps}(c) shows the macro-particle density (dashed blue line) and the fluid density (red line).  There are no adjustable parameters in this comparison; the respective models used
identical numerical parameters.  Clearly the agreement is remarkable.  The macro-particle model has virtually no noise (in part due to the quartic $\rho_k$), even in the density.  No smoothing or
filtering of any kind has been applied to the macro-particle results.  Note also, as in the examples of Section~\ref{Lagrangian-PIC}, there are no signs of grid-heating.

\section{Conclusions}

From a discretized Lagrangian, we have derived a time-explicit, energy-conserving algorithm for modeling relativistic electromagnetic kinetic laser-plasma interactions, in the 1-$\tfrac12$ dimensional
case.  Realizations of this algorithm were developed in the lab frame using both a fourth order Runge--Kutta method and a split-step second order Runge--Kutta/Crank--Nicolson method to integrate the
system in time.  We have shown that with both integrators and for two different physical scenarios the error in energy conservation depends only on temporal discretization, as expected from a
discretized Noether's theorem.  A further advantage of the method was illustrated in its flexibility to accommodate coordinate transformation by extending the formulation to moving window coordinates.
Finally, all of the examples presented showed a reduction of numerical noise as compared to what would be expected from the standard PIC algorithm.  The Lagrangian formulation naturally leads to the
possibility of a (canonical) Hamiltonian formulation and thus the prospect of using a symplectic integrator for both the macro-particles and fields.  A symplectic integrator has been demonstrated for
the electrostatic case with promising computational performance \cite{Shadwick:2014aa}.  The electromagnetic case leads to a significant complication as the kinetic energy depends on both coordinates
and momenta and thus the usual splitting approach fails; this is under active investigation by the authors and will be reported on in a subsequent publication.

\section*{Acknowledgments}

BAS would like to acknowledge helpful conversations with J.~Paxon Reyes, John M. Finn, Michael Carri\'e, and David L.~Bruhwiler.

\bibliography{em1d,notes}
\vfill
\newpage

\begin{IEEEbiography}[{\includegraphics[width=1in,height=1.25in,clip,keepaspectratio]{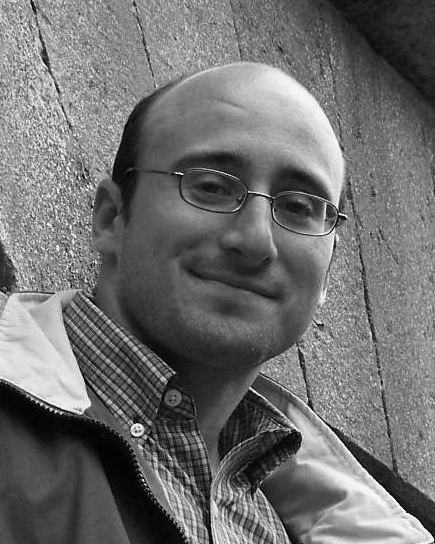}}]{A.\ B.\ Stamm}%
Alexander Stamm is a GAANN fellow working on his Ph.D.\ in Physics at the University of Nebraska-Lincoln, where he received his M.S.\ degree in Physics in 2011.  Prior to this he completed an M.~Eng.\
degree in Material Science and Engineering in 2009 and a B.S.\ degree in Electrical and Computer Engineering in 2007, both at Cornell University.
\end{IEEEbiography}

\begin{IEEEbiography}[{\includegraphics[height=1.25in]{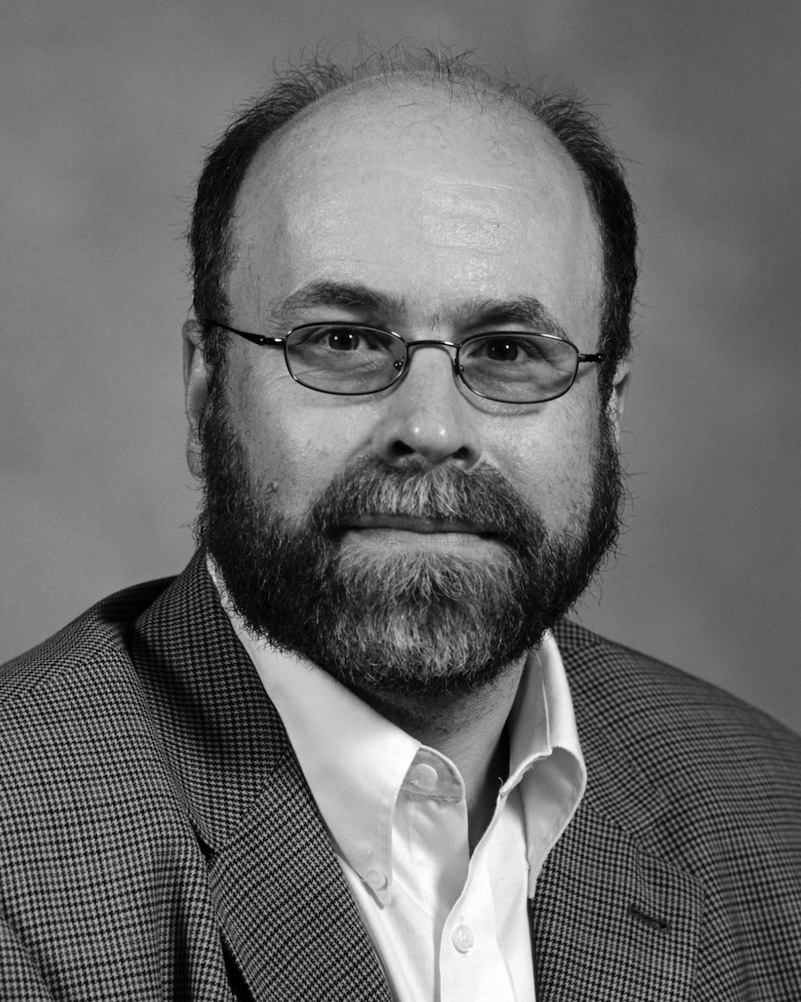}}]{B.~A.\ Shadwick}
Brad Shadwick received his B.Sc. degree in Applied Mathematics from the University of Western Ontario in 1986 and his M.Sc. degree in high-energy physics from the University of Toronto in 1987.  He 
subsequently earned his Ph.D. from The University of Texas at Austin in 1995.  

He was a postdoctoral fellow at the University of California at Berkeley and a scientist at Lawrence Berkeley National 
Laboratory before joining the Department of Physics and Astronomy at the University of Nebraska-Lincoln in 2007 where he is currently an associate professor of physics.  His research interests 
include laser-plasma interactions, plasma-based particle accelerators and computational physics.
\end{IEEEbiography}

\begin{IEEEbiography}[{\includegraphics[height=1.25in]{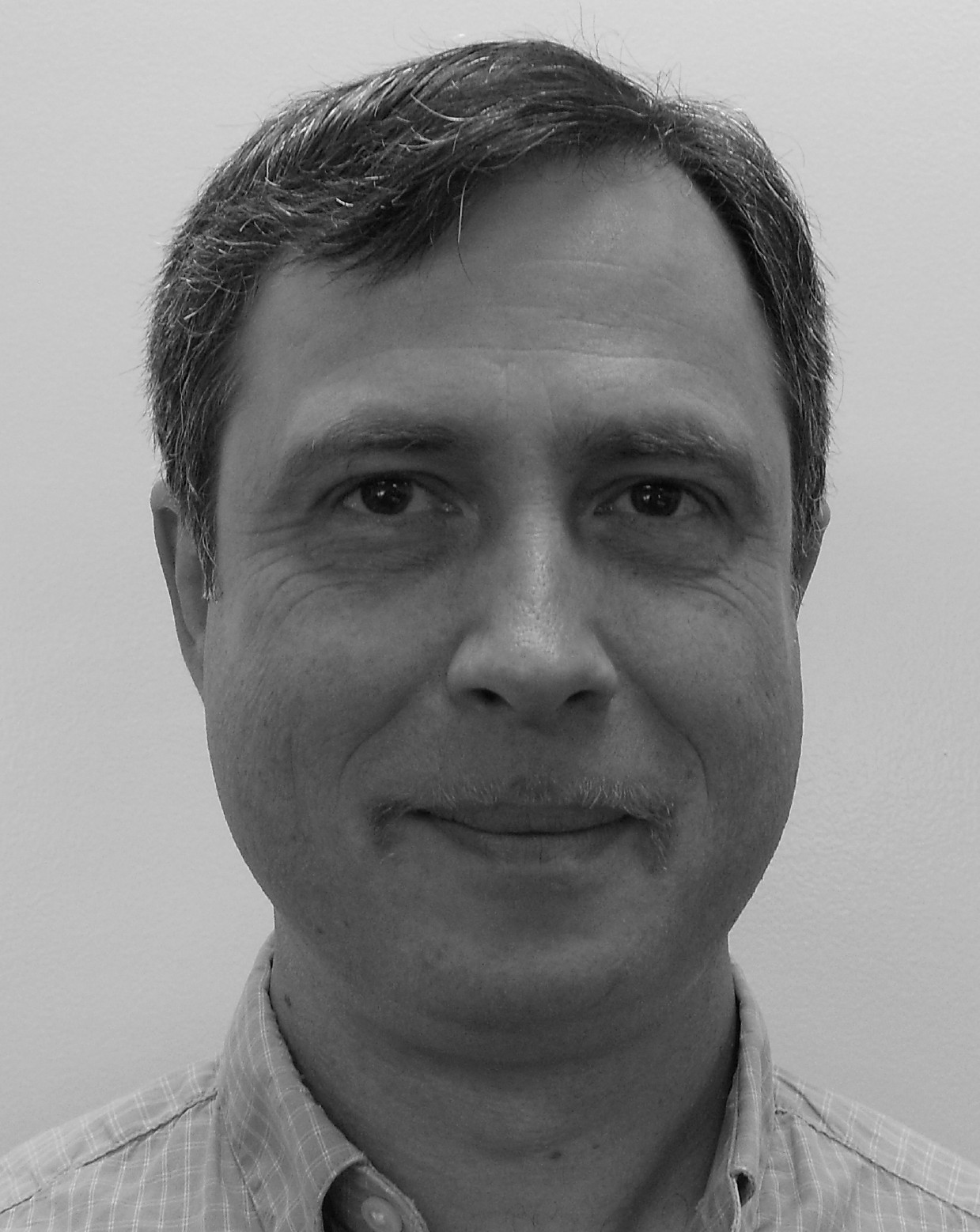}}]{E.~G.\ Evstatiev}
E.~G.\ Evstatiev received his B.S.\ and M.Sc.\ degrees in 1995 in the area of theoretical high-energy and nuclear physics from Sofia University, Sofia, Bulgaria.  He worked at Bulgarian Academy of
Sciences from 1996 to 1997 prior to receiving his Ph.D.\ degree from the University of Texas at Austin in 2004.  He was a postdoctoral associate at the Los Alamos National Laboratory from 2005 to 2008 and at
the University of Nebraska-Lincoln from 2008 to 2009.  He is currently a staff scientist at FAR-TECH, Inc., San Diego, CA. His research includes computational plasma physics, laser-plasma
interactions, magnetohydrodynamics, inertial electrostatic confinement, fluid dynamics, non-linear dynamics, completely integrable dynamical systems and solitons.

\end{IEEEbiography}
\vfill

\end{document}